\documentclass[twocolumn,pra,showpacs]{revtex4}

\usepackage{graphicx}
\usepackage{rotating}
\usepackage{amsmath}
\usepackage{amsfonts}
\usepackage{amssymb}
\usepackage{enumerate}
\usepackage{longtable}
\usepackage{dcolumn}% Align table columns on decimal point
\usepackage{bm}

\begin{document}

\title{Exact Results on Dynamical Decoupling by 
$\pi$ Pulses in Quantum Information Processes}

\author{G\"otz S. Uhrig}
\email{goetz.uhrig@tu-dortmund.de}
\affiliation{Chair for Theoretical Physics I,
 Dortmund University of Technology,
 Otto-Hahn Stra\ss{}e 4, 44221 Dortmund, Germany}

\date{\rm\today}

\begin{abstract}
The aim of dynamical decoupling consists in the suppression of decoherence
by appropriate coherent control of a quantum register. Effectively, the
interaction with the environment is reduced.
In particular, a sequence of $\pi$ pulses is considered. 
Here we present exact results on the suppression of the coupling
of a quantum bit to its environment by optimized sequences of $\pi$ pulses.
The effect of various cutoffs of the spectral density of the
environment is investigated. As a result we show that the harder the cutoff
is the better an optimized pulse sequence can deal with it.
For cutoffs which are neither completely hard nor very soft we advocate
iterated optimized sequences.
\end{abstract}

\pacs{03.67.Pp,03.67.Lx,03.65.Yz,03.65.Vf}

%% PACSes
% 03.67.Pp Quantum error correction and other methods for protection against 
% decoherence (see also 03.65.Yz Decoherence; open systems; quantum statistical
%  methods; for decoherence in Bose-Einstein condensates, see 03.75.Gg) 
%  03.67.Lx, Quantum computation,
% 03.65.Yz Decoherence; open systems; quantum statistical methods 
% (see also 03.67.Pp in quantum information; for decoherence in Bose-Einstein 
% condensates, see 03.75.Gg) 
% 03.65.Vf Phases: geometric; dynamic or topological  

\maketitle

\section{Introduction}

% Why
Almost six decades ago in 1950 Hahn demonstrated that spin echos in liquid 
NMR can be obtained by applying a $\pi$ pulse in the middle of a time 
interval \cite{hahn50}. This idea was developed further by Carr and Purcell who
proposed iterated cycles of two $\pi$ pulses to reduce the effect
of unwanted interactions \cite{carr54}. Further refinements
were introduced by Meiboom and Gill \cite{meibo58}. Since then this technique
of coherent control has been established in NMR, see e.g.\ Ref.\
\onlinecite{haebe76}.

The fascinating possibilities of quantum information have stimulated a great
interest in the coherent control of small quantum systems, see e.g.\
Ref.\ \onlinecite{zolle05}. The idea to preserve coherence by iterated
$\pi$ pulses periodic in time was rediscovered in the context of quantum 
information by Viola and Lloyd \cite{viola98} and by Ban \cite{ban98} in 1998
for a spin-boson model and subsequently generalized to open systems 
\cite{viola99a}; a short review is found in Ref.\ \onlinecite{viola02}.
For symmetry groups with inefficient representations
randomized protocols are advocated, see
Ref.\ \onlinecite{santo08} and references therein.

Recently, periodically iterated Carr-Purcell cycles have been advocated
for the preservation of the coherence of the electronic spin in quantum
dots \cite{yao07,witze07a}. Besides periodic iteration of pulse cycles
also concatenations of cycles were proposed 
and it was shown that they suppress decoherence in higher orders $t^l$ in the
length of the time interval $t$ \cite{khodj05,khodj07,yao07,witze07c}. But the
achieved exponent $l$ grows only logarithmically in the number of pulses
$n$.

In parallel, the author showed that neither the iteration nor the concatenation
of the Carr-Purcell  two-pulse cycle is the optimum strategy
for a single-axis bosonic bath.
\cite{uhrig07}. Cycles with $n$ pulses at the instants $\delta_j t$ 
\begin{equation}
\label{eq:opt-seq}
\delta_j = \sin^2\left[\pi j/(2n+2) \right]
\end{equation}
achieve the optimum suppression of decoherence in the sense that any deviation
of the signal occurs with a high power in $t$, namely $t^{2n+2}$ where
$n$ is the number of pulses, i.e.\ only a linear effort is required.
The Carr-Purcell cycle is retrieved for $n=2$. Up to $n\leq 5$ the
result \eqref{eq:opt-seq}  was previously shown for general models 
\cite{dhar06}.

Then, Lee et al.\ \cite{lee08a} observed in numerical simulations of spin baths
that the pulse sequences obeying Eq.\ \eqref{eq:opt-seq} suppress
also the decoherence for spin baths. They could also show analytically
that the  sequence defined by \eqref{eq:opt-seq} works for the most  general
phase decoherence model up to $n=9$. We have been able to extend this
analytical proof up to $n\leq 14$.
An unrestricted derivation, however, is still lacking.
Lee et al.\ \cite{lee08a} also argued that the optimized sequence
\eqref{eq:opt-seq} works well only when the expansion in time is applicable.

%Experiment
There is a multitude of experimental results in NMR on coherent control
and the suppression of decoherence, see e.g.\ Ref.\ \onlinecite{haebe76}.
We highlight results in the context of quantum information related to
pulse sequences \cite{kroja06}; for a overview see Ref.\ \onlinecite{vande04}.
But also in semi-conductor physics there are many encouraging results in 
prolonging the  coherence time of a qubit by $\pi$ pulses
\cite{frava05,petta05,morto06,greil06a,greil06b}. 
In experiment, one must trade off
between the advantages of the suppression of decoherence by multiply
applied pulses with the detrimental effects of imperfect realizations of
pulses, for instance the finite duration of a pulse so that it
cannot be regarded as instantaneous \cite{pasin08a}.

%Aim
The aim of the present article is threefold. First, we provide the explicit
calculations leading to the important relation \eqref{eq:opt-seq}. 
Second, we generalize the previous result \cite{uhrig07} on a particular signal
to a statement on the unitary time evolution. Thereby, we provide the
general proof for the applicability of \eqref{eq:opt-seq} for 
an arbitrary initial quantum state.
Third, we use various spectral densities $J(\omega)$ in the 
spin-boson model to discuss
under which conditions the optimized sequence works well, namely when the 
high-energy cutoff of the decohering environment is hard enough. 
To cope with medium hard cutoffs we propose iterated sequences
of short optimized cycles of pulses. 

% Aufbau
The article's setup is as follows. In the following Sect.\ \ref{sec:sb-model} 
the explicit calculation for the spin-boson model is presented, both
for the signal in a generic decoherence experimenet 
and for the general time evolution.
The results are also given for classical noise.
The subsequent Sect.\ \ref{sec:quant-sys} 
treats the general phase decoherence model.
Section \ref{sec:uv} presents a discussion of the applicability
of the optimized sequences and establishes a link to the nature of
the high energy cutoff. The conclusions \ref{sec:conclusio} 
summarize the results and discuss their implications for further
developments.

\section{Spin-Boson Model}
\label{sec:sb-model}

We consider the model given by the Hamilton operator
\begin{equation}
\label{eq:hamilton}
H=\sum_i\omega_i b_i^{\dagger}b_i^{\phantom\dagger}+\frac{1}{2}
\sigma_z\sum_i \lambda_i(b_i^{\dagger} + b_i^{\phantom\dagger}) + E
\end{equation}
consisting of a single qubit interpreted as spin $S=1/2$,
whose operators are the Pauli matrices $\sigma_x, \sigma_y$, and
$\sigma_z$. The environment is given by  the bosonic bath with annihilation 
(creation) operators $b_i^{(\dagger)}$. 
The constant $E$ sets the energy offset. The properties
of the bath are defined by the set of real parameters $\{\lambda_i,\omega_i\}$.
This information is conveniently encoded in the spectral density 
\cite{legge87,weiss99}
\begin{equation}
\label{eq:specfunc}
J(\omega) = \sum_i \lambda_i^2 \delta(\omega-\omega_i).
\end{equation}
Obviously, $H$ in \eqref{eq:hamilton} does not allow for spin flips
since it commutes with $\sigma_z$. Physically this means that
the decay time $T_1$ of a magnetization along
$z$  is infinite. But the decoherence of a magnetization in
the $\sigma_x\sigma_y$-plane is captured by $H$ so that the decay time $T_2$
can be investigated in the framework of this model.

The Hamiltonian $H$ in \eqref{eq:hamilton} is analytically 
diagonalizable. For any operator $A$ we will use the notation
\begin{equation}
\label{eq:eff-op}
A^\text{eff}:= U A U^\dagger.
\end{equation}
The unitary transformation $U$ is chosen so that it diagonalizes
$H$
\begin{equation}
\label{eq:diag}
H^\text{eff} =\sum_i\omega_i b_i^{\dagger}b_i^{\phantom\dagger}+ \Delta E\ .
\end{equation}
The appropriate unitary transformation is
\begin{equation}
\label{eq:unitary}
U= \exp(\sigma_z  K).
\end{equation}
The operator $K$ is anti-Hermitean
\begin{equation}
\label{eq:generator1}
K=\sum_i\frac{\lambda_i}{2\omega_i}(b_i^{\dagger}-b_i^{\phantom\dagger})
\end{equation}
so that $U$ is indeed unitary. The energy  offset
after the transformation reads
\begin{equation}
\Delta E = E-\int_0^\infty \frac{J(\omega)}{\omega}d\omega.
\end{equation}
But the global energy offset is not measurable, so that its
quantitative form does not matter.

\subsection{Signal without $\pi$ Pulses}

Here we discuss the simple experimental setup
without any $\pi$ pulses. We start from the
state $|\uparrow\rangle$. Then a $\pi/2$ pulse is applied
to rotate the spin from the $z$-direction to 
the $xy$-plane. To be specific, we rotate the spin
about $x$ by the angle $\gamma$ with the help of the
unitary transformation
\begin{subequations}
\begin{eqnarray}
D_x(\gamma) &:=& \exp(-i\gamma\sigma_x/2)\\
& = & \cos(\gamma/2) +i\sigma_x \sin(\gamma/2).
\end{eqnarray}
\end{subequations}
The rotation is best seen by stating that
\begin{equation}
D_x(\gamma)^\dagger \sigma_z D_x(\gamma) =  \sigma_z\cos\gamma +
\sigma_y \sin\gamma.
\end{equation}
For $\gamma=\pi/2$ a spin along $z$ is turned into a spin along $y$.
We will use $D_x(\pi/2)=(1+i\sigma_x)/\sqrt{2}$.

In the $xy$ plane the spin will evolve. After the time $t$ a measurement
of $\sigma_y$ yields the signal
\begin{equation}
s(t) = \langle \uparrow | D_x(\pi/2)^\dagger  \exp(iHt) \sigma_y  
\exp(-iHt)  D_x(\pi/2)| \uparrow \rangle. 
\end{equation}
Since $H$ does not induce spin flips and 
$\langle \uparrow |\sigma_y |\uparrow \rangle=0 
=\langle \downarrow |\sigma_y |\downarrow \rangle$ we know that
$\langle \uparrow |  \exp(iHt) \sigma_y  \exp(-iHt) |\uparrow \rangle=0$ and 
$\langle \uparrow |\sigma_x  \exp(iHt) \sigma_y  \exp(-iHt) \sigma_x
|\uparrow \rangle=0$. Hence the signal is given by
\begin{equation}
\label{eq:s-im1}
s(t) = \mathfrak{Im} \langle \uparrow |\sigma_x \exp(iHt) \sigma_y  
\exp(-iHt) | \uparrow \rangle.
\end{equation}

To evaluate this expression explicitly we change to the basis in
which $H$ is diagonal
\begin{equation}
\label{eq:s-diag}
s(t) = \mathfrak{Im} \langle\uparrow|\sigma_x^\text{eff} 
\exp(iH^\text{eff}t) \sigma_y^\text{eff} 
\exp(-iH^\text{eff}t) |\uparrow\rangle.
\end{equation}
Note that the state $|\uparrow \rangle$ is not altered by $U$.
For the explicit calculation of the effective operators we use 
\begin{equation}
\label{eq:pauli1}
\sigma_{x/y}\sigma_z = -\sigma_z\sigma_{x/y}
\end{equation}
 and obtain
\begin{subequations}
\begin{eqnarray}
\sigma_{x/y}^\text{eff} &=& \exp(\sigma_z  K)\sigma_{x/y} \exp(-\sigma_z  K)\\
&=& \exp(2\sigma_z  K)\sigma_{x/y}.
\label{eq:sigma-eff}
\end{eqnarray}
\end{subequations}
Hence the action on particular spin states is
\begin{subequations}
\label{eq:spinids}
\begin{eqnarray}
\sigma_x^\text{eff} |\uparrow/\downarrow\rangle &=& 
\exp(\mp 2K)|\downarrow/\uparrow\rangle\\
\sigma_y^\text{eff} |\uparrow/\downarrow\rangle &=& 
\pm i\exp(\mp 2K)|\downarrow/\uparrow\rangle,
\end{eqnarray}
\end{subequations}
where either the first spin orientation and the upper sign holds
or the second spin orientation and the lower sign.

Turning to the time dependence we define generally
the time dependent operators
\begin{equation}
\label{eq:time-dependence}
A(t):= \exp(iH^\text{eff}t) A \exp(-iH^\text{eff}t).
\end{equation}
Note that $H^\text{eff}$ contains only the bosonic 
degrees of freedom and it is diagonal. Hence it is easy to see that
\begin{subequations}
\begin{eqnarray}
b^\dagger_i(t) &=& b^\dagger_i\exp(i\omega_it)\\
b^{\phantom\dagger}_i(t) &=& b^{\phantom\dagger}_i\exp(-i\omega_it)
\end{eqnarray}
\end{subequations}
whence
\begin{equation}
K(t)= \sum_i\frac{\lambda_i}{2\omega_i}(b_i^{\dagger}\exp(i\omega_it)
-b_i^{\phantom\dagger}\exp(-i\omega_it)).
\end{equation}
With these definitions the identities
\eqref{eq:spinids} apply also to the time dependent 
operators $\sigma^\text{eff}(t)$ and $K(t)$
\begin{subequations}
\label{eq:spinids-time}
\begin{eqnarray}
\sigma_x^\text{eff}(t) |\uparrow/\downarrow\rangle &=& 
\exp(\mp 2K(t))|\downarrow/\uparrow\rangle\\
\sigma_y^\text{eff}(t) |\uparrow/\downarrow\rangle &=& 
\pm i\exp(\mp 2K(t))|\downarrow/\uparrow\rangle.
\label{eq:spinids-time-b}
\end{eqnarray}
\end{subequations}

With these identities we can write for the signal 
\begin{subequations}
\begin{eqnarray}
\label{eq:s-res1}
s(t) &=& \mathfrak{Im} \langle\uparrow|\sigma_x^\text{eff}(0) 
 \sigma_y^\text{eff}(t)  |\uparrow\rangle\\
&=& \mathfrak{Im}\, i\langle\downarrow| \exp( 2K(0))
\exp(- 2K(t))  |\downarrow\rangle\qquad \\
&=& \mathfrak{Re} \langle \exp( 2K(0))
\exp(- 2K(t))\rangle,
\label{eq:s-res1-c}
\end{eqnarray}
\end{subequations}
where we took the expectation value in the spin sector in \eqref{eq:s-res1-c}
so that only a bosonic expectation value with respect to the bilinear
$H^\text{eff}$ must be computed. This is eased by the
Baker-Campbell-Hausdorff (BCH) formula 
\begin{equation}
\label{eq:bch}
\exp(A)\exp(B)=\exp(A+B)\exp([A,B]/2)
\end{equation}
which is valid if $[A,B]$ commutes with $A$ and $B$. This yields
\begin{equation}
s(t) = \mathfrak{Re}\exp(-2[K(0),K(t)])\langle \exp(- 2\Delta K)\rangle
\end{equation}
with $\Delta K:=K(t)-K(0)$. Any expectation value of 
an exponential of a linear bosonic operator $A$ with respect to a bilinear
Hamiltonian such as $H^\text{eff}$
can be reduced to the exponential of an expectation value
by 
\begin{equation}
\label{eq:expect}
\langle\exp(A)\rangle =\exp(\langle A^2\rangle/2).
\end{equation}
 Hence we have
\begin{equation}
s(t) = \mathfrak{Re}\exp(-2[K(0),K(t)])
\exp(2\langle \Delta K^2\rangle)
\end{equation}
which simplifies due to the Hermitecity of $\Delta K^2$ to
\begin{equation}
\label{eq:qm-result1}
s(t) = \cos(2\varphi(t)) \exp(-2\chi(t))
\end{equation}
where the phase is given by
\begin{subequations}
\begin{eqnarray}
\varphi(t) &:=& i[K(0),K(t)]\\
&=& -i\sum_i \frac{\lambda_i^2}{4\omega_i^2}(e^{i\omega_it}
-e^{-i\omega_it})\\
&=& \sum_i \frac{\lambda_i^2}{2\omega_i^2} \sin(\omega_it)\\
&=&\frac{1}{2} \int_0^\infty \frac{J(\omega)}{\omega^2} \sin(\omega t)d\omega.
\end{eqnarray}
\end{subequations}
The exponential suppression is given by
\begin{equation}
\chi(t)  := -\langle \Delta K^2\rangle
\end{equation}where
\begin{equation}
\Delta K =  \sum_i\frac{\lambda_i}{2\omega_i}
\left[b_i^{\dagger}(e^{i\omega_it}
-1)-b_i^{\phantom\dagger}(e^{-i\omega_it}-1)\right]
\end{equation}
whence we obtain
\begin{equation}
\chi(t)  = \sum_i\frac{\lambda_i^2}{4\omega_i^2} 4\sin^2(\omega_i t/2)
\langle b_i^{\dagger}b_i^{\phantom\dagger}+  
b_i^{\phantom\dagger} b_i^{\dagger}\rangle.
\end{equation}
The bosonic occupation is such that the last expectation value
equals $\coth(\beta\omega_i/2)$ so that we finally have
\begin{equation}
\label{eq:qm-result2}
\chi(t)  = \int_0^\infty J(\omega)\frac{\sin^2(\omega t/2)}{\omega^2}
\coth(\beta\omega/2)d\omega.
\end{equation}
This concludes the derivation of the signal without
any dynamical decoupling. The formulae (6) in Ref.\ \onlinecite{uhrig07}
are rederived in all detail. The above derivation sets the stage
for the derivation in the case of dynamical decoupling by sequences
of $\pi$ pulses.

\subsection{Signal with $\pi$ Pulses}

Here we consider a sequence of $\pi$ pulses which are applied
at the instants of time $\delta_i t$ with $i\in \{1,2,\ldots,n\}$
so that $n$ pulses are applied and the total time interval $t$ is
divided into $n+1$ subintervals. For notational convenience
we set $\delta_0=0$ and $\delta_{n+1}=1.$ It is understood that 
$\delta_{i+1}>\delta_i$ for all  $i\in \{0,1,2,\ldots,n\}$.

The $\pi$ pulses are taken to be ideal, that means they
are instantaneous so that during their application no coupling
to the bath needs to be considered. The possible workarounds
if this is not justified in experiment are discussed elsewhere \cite{pasin08a}.
For simplicity, we take the $\pi$ pulses to be realized as 
rotations about $\sigma_y$
\begin{subequations}
\begin{eqnarray}
D_y(\gamma) &:=& \exp(-i\gamma\sigma_y/2)\\
& = & \cos(\gamma/2) +i\sigma_y \sin(\gamma/2).
\end{eqnarray}
\end{subequations}
which implies for $\gamma=\pi$ simply $D_y(\pi)=i\sigma_y$.
Below we will use $\sigma_y$ only because the factor $i$ 
corresponds to an irrelevant global phase shift.

The signal is given in general as before by
\begin{equation}
\label{eq:s-general1}
s(t) = \mathfrak{Im} \langle \uparrow |\sigma_x \widetilde{R}^\dagger
\sigma_y   \widetilde{R} | \uparrow \rangle
\end{equation}
where the time evolution is changed its form $\exp(-iHt)$ in \eqref{eq:s-im1}
  to 
\begin{eqnarray}
\nonumber
\widetilde{R} &:=& e^{-iH(\delta_{n+1}-\delta_n)t}\sigma_y
e^{-iH(\delta_{n}-\delta_{n-1})t}\sigma_y
\ldots\\
&&\ldots  e^{-iH(\delta_{2}-\delta_{1})t} \sigma_y
e^{-iH(\delta_{1}-\delta_{0})t}.
\label{eq:tildeR}
\end{eqnarray}
The expression \eqref{eq:tildeR}  becomes much more compact if written after
the diagonalisation by $U$ as given in \eqref{eq:unitary}
\begin{equation}
\label{eq:Reff}
\widetilde{R}^\text{eff} = e^{-iH^\text{eff}t} R^\text{eff}
\end{equation}
where we express $R^\text{eff}$ based on 
\eqref{eq:time-dependence} in the form
\begin{equation}
\label{eq:Reff-explicit}
R^\text{eff}=
\sigma^\text{eff}_y(\delta_nt)\;
\sigma^\text{eff}_y(\delta_{n-1}t)\;\ldots\;
\sigma^\text{eff}_y(\delta_2t)\;
\sigma^\text{eff}_y(\delta_1t).
\end{equation}
Then we arrive easily at 
\begin{equation}
\label{eq:s-general2}
s(t) = \mathfrak{Im} \langle \uparrow |\sigma_x^\text{eff}(0) 
R^{\text{eff}\;\dagger}
\sigma_y^\text{eff}(t)   R^\text{eff} | \uparrow \rangle.
\end{equation}

Equation \eqref{eq:s-general2} can be converted by means of
\eqref{eq:spinids-time} into the following purely bosonic expression
\begin{eqnarray}
\nonumber
s(t) &=&\mathfrak{Im}\Big\langle
e^{2K(0)}\;ie^{-2K(\delta_1t)}\;(-i)e^{2K(\delta_2t)}\;\ldots\\
\nonumber
&& \ldots\; (-i)(-1)^n e^{(-1)^n 2K(\delta_nt)}
\\ \nonumber &&
(-i)(-1)^{n+1} e^{(-1)^{n+1}2K(t)}\\
\nonumber
&&(-i)(-1)^n e^{(-1)^n2K(\delta_nt)}\;\ldots \\
&&
\ldots\;(-i)e^{2K(\delta_2t)} \;ie^{-2K(\delta_1t)}\Big\rangle.
\end{eqnarray}
Counting the factors $(-1)$ and $i$ one finds that they all combine to 
a single factor $i$. This is easiest seen by combining the prefactors
in front of each of the terms $(-i)(-1)^n e^{(-1)^n 2K(\delta_nt)}$
which all occur twice so that each kind of these terms provides a factor
$(-i)^2=-1$ yielding a total factor $(-1)^n$. This is multiplied
with $(-i)(-1)^{n+1}$ from the prefactor of $e^{(-1)^n2K(\delta_nt)}$
which is the only term occuring only once. Hence we have
\begin{eqnarray}
\nonumber
s(t) &=&\mathfrak{Re}\Big\langle
e^{2K(0)}\;e^{-2K(\delta_1t)}\;e^{2K(\delta_2t)}\;\ldots\\
\nonumber
&\ldots&\!\! e^{(-1)^n 2K(\delta_nt)}\;
e^{(-1)^{n+1}2K(t)}\;
e^{(-1)^n2K(\delta_nt)}\ldots\\
&&
\ldots\;e^{2K(\delta_2t)} \;e^{-2K(\delta_1t)}\Big\rangle.
\label{eq:s-general4}
\end{eqnarray}
Applying the BCH formula \eqref{eq:bch} yields
\begin{subequations}
\label{eq:s-general3}
\begin{eqnarray}
\label{eq:s-general3a}
s(t) &=& \mathfrak{Re} \exp(2i\varphi_n(t))
\left\langle \exp(2\Delta_n K)\right\rangle\\
\label{eq:finres}
 &=& \cos(2\varphi_n(t))
\exp\left(-2\chi_n(t)\right)
\end{eqnarray}
\end{subequations}
where we used the identity \eqref{eq:expect} to obtain
the second line. Therein the suppression 
$\chi_n(t):=-\langle \Delta_n K^2\rangle$
results from
\begin{subequations}
\label{eq:multi-delta}
\begin{eqnarray}
\Delta_n K \! &:=& K(0)\!+\! (-1)^{n+1}K(t) \!+\!
2\!\sum_{i=1}^n (-1)^n K(\delta_i t)\qquad\\
&=&
\sum_i\frac{\lambda_i}{2\omega_i}(b_i^{\dagger}\;y_n(\omega_i t)-
b_i^{\phantom\dagger}\;y_n^*(\omega_i t)),
\label{eq:suppression-general-b}
\end{eqnarray}
\end{subequations}
where
\begin{equation}
\label{eq:def-yn}
y_n(z) :=  1 + (-1)^{n+1} e^{iz} + 2\sum_{j=1}^n(-1)^j e^{iz\delta_j} .
\end{equation}
Thereby we arrive at
\begin{subequations}
\label{eq:finchi}
\begin{eqnarray}
\chi_n(t) &=& \sum_i\frac{\lambda_i^2}{4\omega_i^2} 
\left| y_n(\omega_i t)\right|^2
\langle b_i^{\dagger}b_i^{\phantom\dagger}+  
b_i^{\phantom\dagger} b_i^{\dagger}\rangle\\
\label{eq:finchi-b}
&=& \int_0^\infty J(\omega)\frac{| y_n(\omega t)|^2}{4\omega^2}
\coth(\beta\omega/2)d\omega.
\end{eqnarray}
\end{subequations}
The phase $\varphi_n(t)$ in \eqref{eq:s-general3}
can easily be computed by the following trick. Using \eqref{eq:bch}
we combine the second and third factor in \eqref{eq:s-general4}, i.e., the
two exponentials $e^{-2K(\delta_1t)}\;e^{2K(\delta_2t)}$, 
to one exponential \emph{and} the last and last-but-one factor, i.e.,
$e^{2K(\delta_2t)} \;e^{-2K(\delta_1t)}$. Obviously, the occuring
phases cancel. This procedure can be repeated by including the
factor $e^{-2K(\delta_3t)}$ next both in the growing last exponential
and in the second exponential. Iteration up to and including the factor
$e^{(-1)^{n+1}2K(\delta_{n+1}t)}$, which can be thought as being
split into $e^{(-1)^{n+1}2K(\delta_{n+1}t)/2}\;
e^{(-1)^{n+1}2K(\delta_{n+1}t)/2}$, leads to two exponentials whose
respective arguments contain all term $K(\delta_j t)$ except the very first
$K(0)$. Furthermore, the two respective arguments are equal so that
the exponentials can be combined without further phase yielding
\begin{equation}
s(t)=\mathfrak{Re}\left\langle e^{2K(0)}
e^{2\Delta_nK -2K(0)}\right\rangle.
\end{equation}
From this equation we arrive at \eqref{eq:s-general3a}
by defining
\begin{subequations}
\label{eq:finphi}
\begin{eqnarray}
\label{eq:phi-general-final-a}
\varphi_n(t) &:=& -i[K(0),\Delta_nK]\\
&=& i \sum_i \frac{\lambda_i^2}{4\omega_i^2}(y_n(\omega_i t)
- y_n^*(\omega_i t))
\label{eq:phi-general-final-b}
\\
&=& i \int_0^\infty \frac{J(\omega)}{4\omega^2} (y_n(\omega t)
- y_n^*(\omega t)) d\omega\\
&=& \int_0^\infty \frac{J(\omega)}{2\omega^2} x_n(\omega t)d\omega
\label{eq:phi-general-final-d}
\end{eqnarray}
\end{subequations}
where we used \eqref{eq:suppression-general-b} in the second line
\eqref{eq:phi-general-final-b}. The last line \eqref{eq:phi-general-final-d}
reproduces Eq.\ (8b) in Ref.\ \onlinecite{uhrig07} with
\begin{subequations}
\label{eq:x-from-y}
\begin{eqnarray}
\label{eq:x-from-y-a}
x_n(z) &:=& i(y_n(z) - y_n^*(z))/2\\
&=& -\Im  y_n(z) \\
&=& (-1)^n\sin(z) + 2\sum_{j=1}^n(-1)^{j+1}\sin(z\delta_j),\qquad
\label{eq:x-from-y-b}
\end{eqnarray}
\end{subequations}
where the last line \eqref{eq:x-from-y-b}
corrects Eq.\ (9) in Ref.\ \onlinecite{uhrig07} in the factor 2
in front of the sum.

Thereby, we have derived all the results used in the analysis in the
previous paper \cite{uhrig07}.

\subsection{Optimization of the Sequence of $\pi$ Pulses}
\label{ss:optimization}

A particular asset of the equations
(\ref{eq:finres},\ref{eq:finchi},\ref{eq:finphi}) together with 
\eqref{eq:x-from-y-a} is that it is obvious that any deviation of
the signal $s(t)$ from unity is kept as low as possible if 
$|y_n(z)|$ is kept as small as possible. Note that this strategy 
holds equally well for $\varphi_n$ and for $\chi_n$. If $y_n$ is of the
order $p$ in some small parameter $p$, for instance $p=t^{n+1}$,
then $\chi_n={\cal O}(p^2)$ whence we deduce $\exp(-2\chi_n)=1-{\cal O}(p^2)$.
{} In analogy, we find $\varphi_n={\cal O}(p)$ whence we deduce 
$\cos(-2\varphi_n)=1-{\cal O}(p^2)$ so that both factors are close to 
unity in the same way. Hence the total signal $s(t)$ is close
to unity in this order $s(t)=1-{\cal O}(p^2)$.

So our aim is to choose the $n$ instants $\{\delta_j\}$ such that
$y_n(z)$ is as small as possible. The best way to do so is to make the
first $n$ derivatives of $y_n(z)$ vanish. Note that $y_n(0)=0$
for any sequence $\{\delta_j\}$. The $m$th derivative
reads ($m>0$)
\begin{equation}
\partial_z^m y_n\big|_{z=0}
= i^m\Big( (-1)^{n+1} + 2\sum_{j=1}^n(-1)^j \delta_j^m  \Big).
\end{equation}
Hence we have to solve the set of nonlinear equations
\begin{equation}
\label{eq:condition}
0=  (-1)^{n+1} + 2\sum_{j=1}^n(-1)^j \delta_j^m 
\end{equation}
for $m\in\{1,2,\ldots,n\}$. For finite $n$, solutions can easily
be found analytically \cite{dhar06} and numerically. 
Closer inspection of these numerical solutions reveals
that they are excellently described by the condition \eqref{eq:opt-seq}.

Indeed, we can prove that \eqref{eq:opt-seq} is a valid solution for the
set of equations \eqref{eq:condition}. To do so we choose a little detour
by considering 
\begin{equation}
\widetilde{y}_n(h)\big|_{h=z/2}:=\exp(-iz/2){y}_n(z).
\end{equation}
Obviously the equivalence 
\begin{equation}
{y}_n(z)={\cal O}(z^{n+1}) \Leftrightarrow 
\widetilde{y}_n(h)={\cal O}(h^{n+1})
\end{equation}
 holds so that the vanishing of the first $n$ derivatives of $y_n(z)$ is
equivalent to the vanishing of the first $n$ derivatives of 
$\widetilde{y}_n(h)$. The choice \eqref{eq:opt-seq} implies by
standard trigonometric identities
\begin{equation}
\delta_j = 1/2-\cos(\pi j/(n+1))/2.
\end{equation}
Inserting this choice into $\widetilde{y}_n(t)$ yields
\begin{subequations}
\begin{eqnarray}
\nonumber
\widetilde{y}_n(h) &=& e^{-ih} +(-1)^{n+1}e^{ih}
+\\
&& + 2\sum_{j=1}^n(-1)^j \exp[-ih\cos(\pi j/(n+1))]\\
&=& \sum_{j=-n-1}^n(-1)^j \exp[-ih\cos(\pi j/(n+1))].\qquad\
\end{eqnarray}
\end{subequations}
Obviously $\widetilde{y}_n(0)=0$. The $m$th derivative ($m>0$) reads
\begin{equation}
\partial_h^m \widetilde{y}_n\big|_{h=0}
= (-i)^m \sum_{j=-n-1}^n(-1)^j \cos^m(\pi j/(n+1)).
\end{equation}
We compute explicitly $d_m:=(2i)^m \partial_h^m \widetilde{y}_n\big|_{h=0}$
\begin{subequations}
\begin{eqnarray}
d_m &=&\sum_{j=-n-1}^n(-1)^j [e^{i\pi j/(n+1)} - e^{-i\pi j/(n+1)})^m
\\
&=&\sum_{\nu=0}^m\left(
\begin{array}{c}
m\\
\nu
\end{array} \right)
\sum_{j=-n-1}^n \exp\left(\frac{i\pi j(2\nu-m+1)}{n+1}\right).\qquad
\end{eqnarray}
\end{subequations}
The last sum, however, vanishes for $m< n+1$
\begin{subequations}
\begin{eqnarray}
\nonumber
&& \sum_{j=-n-1}^n \exp\left(\frac{i\pi j (2\nu-m+1)}{n+1}\right)
= \qquad \qquad \qquad\qquad  \\
&& \ (-1)^{n+1}\frac{\exp(-i\pi(2\nu-m))-\exp(i\pi(2\nu-m))}
{1+\exp\left((i\pi (2\nu-m))/(n+1)\right)}
\qquad 
\label{eq:denominator}
\\
&& \qquad = 0
\end{eqnarray}
\end{subequations}
since the denominator in \eqref{eq:denominator} remains finite 
in this range. Hence $d_m=0$ and we know $\widetilde{y}_n(h)={\cal O}(h^n)$
and hence ${y}_n(z)={\cal O}(z^n)$. This concludes the formal proof
that \eqref{eq:opt-seq} represents a valid solution of the
set of nonlinear equations \eqref{eq:condition}. We have not presented
a proof that this is the only solution. But we presume that it is
the only one which is physically meaningful with consecutive values
$\delta_{j+1}> \delta_j$.

\subsection{Classical Noise with $\pi$ Pulses}
\label{ss:classical}

In Ref.\ \onlinecite{uhrig07} we argued that the fact that
the optimized sequence \eqref{eq:opt-seq} works independently from
the precise temperature indicates that it applies also to classical,
Gaussian noise. The argument runs qualitatively as follows. 
Because \eqref{eq:opt-seq} is the optimum sequence for all temperatures
it holds of course also for $T\to\infty$. In this limit, the thermal
fluctuations dominate over all the quantum effects and the bath
behaves completely classically.

A crucial corollary is that the pulse sequence can be used for
all kinds of bath at elevated temperatures because all physical
systems behave like classical, Gaussian baths at high temperatures.
Hence the applicability extends beyond the spin-boson model discussed
so far. We will discuss the general validity of \eqref{eq:opt-seq}
in more detail in the next section.

Here we present the calculation for classical noise in order
to establish a quantitative relation. We consider the decoherence
due to 
\begin{equation}
H=f(t)\sigma_z
\end{equation}
where $f(t)$ is a random variable with Gaussian dis\-tri\-bu\-tion. 
\footnote{Note  that we have changed the definition of $f(t)$ by a factor of 2 
relative to our previous work \cite{uhrig07} to keep the notation concise.} 
It is characterized by the expectation values
\begin{subequations}
\begin{eqnarray}
\langle f(t)\rangle &=&0\\
\langle f(t_1)f(t_2)\rangle &=& g(t_1-t_2).
\end{eqnarray}
\end{subequations}
Note the translational invariance in time. Then the signal
$s(t)$  after a $\pi/2$ pulse reads
%\begin{subequations}
\begin{eqnarray}
\nonumber
s(t) &=& \langle \uparrow | D_x(\pi/2)^\dagger  e^{iF(t)\sigma_z}\; 
\sigma_y \; 
e^{-iF(t)\sigma_z}  D_x(\pi/2)| \uparrow \rangle \\
 &=&  \mathfrak{Im} \langle \uparrow |\sigma_x e^{iF(t)\sigma_z} \sigma_y  
e^{-iF(t)\sigma_z} | \uparrow \rangle
\end{eqnarray}
%\end{subequations}
where $F(t):=\int_0^tf(t')dt'$ is the primitive of $f(t)$.
Since $\sigma_{x,y}$ only flip the spin, see for instance
Eq.\ \eqref{eq:spinids} for $K=0$, we may write
\begin{subequations}
\begin{eqnarray}
s(t) &=& \langle e^{-iF(t)} e^{-iF(t)}\rangle\\
 &=& e^{-2 \langle F(t)^2\rangle} 
\label{eq:class-gauss}
\end{eqnarray}
\end{subequations}
where we exploited the properties of Gaussian random variables to
obtain the second line \eqref{eq:class-gauss}. The exponent
can be computed easily
\begin{subequations}
\begin{eqnarray}
\langle F(t)^2\rangle &=& \int_0^t dt_1\int_0^t dt_2 
\langle f(t_1)f(t_2)\rangle\\
\label{eq:g-symmetry}
 &=& 2 \int_0^t dt_1\int_0^{t_1} dt' g(t')\\
 &=& \frac{4}{\pi} 
\int_0^\infty \frac{p(\omega)}{\omega^2}\sin^2(\omega t/2)d\omega.
\label{eq:class-res}
\end{eqnarray}
\end{subequations}
where we used $g(t')=g(-t')$ to obtain \eqref{eq:g-symmetry}
and the Fourier representation for \eqref{eq:class-res}
\begin{equation}
g(t) = \frac{1}{\pi} \int_0^\infty p(\omega) \cos(\omega t) d\omega 
\end{equation}
based on the power spectrum $p(\omega)$. The comparison with the
quantum mechanical result (\ref{eq:qm-result1},\ref{eq:qm-result2}) yields
exactly the same form except that $\varphi(t)=0$ because there are no
operators which might not commute with themselves. The argument
of the exponential $\chi(t)=\langle F(t)^2\rangle$
is identical if we identify 
\begin{equation}
\label{eq:identify} 
(4/\pi) p(\omega) = J(\omega)\coth(\beta\omega/2).
\end{equation}
This provides the quantitative 
correspondence between the classical calculation
and the general quantum mechanical one.

The extension to the signal in presence of the $\pi$ pulses is
also straightforward. The signal is given as in \eqref{eq:s-general1}
except that the time evolution $\widetilde{R}_\text{cl}$ is 
classically given by
\begin{eqnarray}
\nonumber
\widetilde{R}_\text{cl} &=& 
e^{-i\sigma_z\int^{\delta_{n+1}t}_{\delta_n t}f(t)dt}
\sigma_y
e^{-i\sigma_z\int^{\delta_{n}t}_{\delta_{n-1} t}f(t)dt}\sigma_y
\ldots\\
&&\ldots  e^{-i\sigma_z\int^{\delta_{2}t}_{\delta_{1} t}f(t)dt}
\sigma_y
e^{-i\sigma_z\int^{\delta_{1}t}_{\delta_{0} t}f(t)dt}.
\label{eq:tildeRcl}
\end{eqnarray}
Again, the dynamics of the spin is easily computed since it flips 
at each $\sigma_y$ or $\sigma_x$ according to \eqref{eq:spinids}.
The final result is $s(t) =  e^{-2 \langle F_n(t)^2\rangle}$
where $ F_n(t)$ is given by
\begin{equation}
F_n(t) :=\int_{-\infty}^{\infty} f(t') s_n(t') dt'
\end{equation}
where $s_n(t')$ switches the sign according to
\begin{equation}
\label{eq:switch}
s_n(t') := \left\{
\begin{array}{lll}
0  & \text{for} & t' \le 0\\
(-1)^j & \text{for} & \delta_j t< t' \le \delta_{j+1} t \\
0 & \text{for} & t' > t
\end{array}\right.
\end{equation}
for $j\in\{0,1,2,\ldots,n\}$. Note that the Fourier transform $s_n(\omega)$
of $s_n(t')$ is given essentially by $y_n(\omega t)$ 
\begin{equation}
\int_{-\infty}^\infty s_n(t')\exp(i\omega t')dt'
=\frac{i}{\omega}y_n(\omega t).
\end{equation}
Next $\langle F_n(t)^2\rangle$
is expressed as convolution and integral
\begin{subequations}
\begin{eqnarray}
\langle F_n(t)^2\rangle
&=&
\int_{-\infty}^\infty \int_{-\infty}^\infty
dt_1 dt_2 s_n(t_1) g(t_1-t_2)s_n(t_2)\qquad\\
&=& \frac{1}{\pi}
\int_0^\infty |y_n(\omega t)|^2 \frac{p(\omega)}{\omega^2} d\omega.
\end{eqnarray}
\end{subequations}

For the last line, Fourier transformation, Parseval identity and the
symmetry of the integrand are used. Again, we retrieve the 
quantum mechanical result (\ref{eq:finres},\ref{eq:finchi-b})
except for the phase $\varphi_n(t)$ which does not occur at all
in the classical framework. The necessary identification is
the same as before \eqref{eq:identify}.

We conclude that the classical decoherence and the
one due to a quantum bosonic bath coincide except for the phases
if the power spectrum $4p(\omega)/\pi$ is 
identified with the product of spectral density $J(\omega)$ and
bosonic occupation factor $\coth(\beta\omega/2)$. Hence the optimization
of the quantum model applies equally to the classical problem.
Therefore, the optimization \eqref{eq:opt-seq}
applies to all models with (commuting) Gaussian fluctuations.

\subsection{Unitary Time Evolution with $\pi$ Pulses}
\label{ss:unitary-SB}

So far we focused on the signal $s(t)$ as it results from 
a measurement of $\sigma_y$ after a $\pi/2$ pulse around  $\sigma_x$.
This appears to be a special choice. But in view of the
spin rotational symmetry about the $z$ axis it is sufficiently general
to guarantee that the coherence of an arbitrary initial state is preserved
by the optimized pulse sequence. To corroborate this point and to 
prepare for the discussion of the most general model for phase coherence
we discuss the time evolution operator $\widetilde{R}$ of the spin-boson model
in this section.

The unitary operator $\widetilde{R}$ is defined in \eqref{eq:tildeR}. 
Using the identities (\ref{eq:eff-op},\ref{eq:Reff}) we get
\begin{subequations}
\begin{eqnarray}
\widetilde{R} &=& U^\dagger e^{-iH^\text{eff}t} R^\text{eff} U\\
&=& e^{-iH^\text{eff}t} \; R \qquad \text{with}\\
R &=& e^{-\sigma_zK(t)} R^\text{eff} e^{\sigma_zK(0)}.
\end{eqnarray}
\end{subequations}
Inserting the time dependent version of \eqref{eq:sigma-eff} (cf.\
also Eq.\  \eqref{eq:spinids-time-b})
\begin{equation}
\sigma_y^\text{eff}(t) = \exp(2\sigma_z  K(t))\; \sigma_y
\end{equation}
into \eqref{eq:Reff-explicit} and using \eqref{eq:pauli1} yields for $n$ even
\begin{eqnarray}
\nonumber
R\big|_{n\; \text{even}} &=& 
e^{-\sigma_zK(t)}\,e^{2\sigma_zK(\delta_nt)}\,
e^{-2\sigma_zK(\delta_{n-1}t)}\ldots\\
&& \ldots e^{2\sigma_zK(\delta_2t)}\,
e^{-2\sigma_zK(\delta_{1}t)}\,e^{\sigma_zK(0)}
\end{eqnarray}
while for $n$ odd we arrive at
\begin{eqnarray}
\nonumber
R\big|_{n\; \text{odd}} &=& 
\sigma_y\;e^{\sigma_zK(t)}\,e^{-2\sigma_zK(\delta_nt)}\,
e^{2\sigma_zK(\delta_{n-1}t)}\ldots\\
&& \ldots e^{2\sigma_zK(\delta_2t)}\,
e^{-2\sigma_zK(\delta_{1}t)}\,e^{\sigma_zK(0)}.
\end{eqnarray}
These results are combined to yield for the total time evolution
\begin{equation}
\widetilde{R} =
\left\{
\begin{array}{c}
1\\
\sigma_y
\end{array}
\right\}
\exp({-iH^\text{eff}t})\; \exp({-i\varphi_n(t)})\;
\exp({\sigma_z \Delta_nK})
\end{equation}
where the upper entry in the curly brackets refers
to $n$ even, the lower one to $n$ odd.  The multiple difference
is defined and computed in \eqref{eq:multi-delta}. Combining all
the exponents to a single one makes a phase $\phi_n(t)$  occur which can be
computed by commuting the various expression $K(\delta_jt)$ as
required in \eqref{eq:bch}. We do not give the explicit expression
because we do not need it here. What is important is that this phase
is a global one. It is just a real number and it does not depend on the spin; 
no Pauli matrix occurs because $\sigma_z^2=1$. 
Similarly, $H^\text{eff}$ does not depend on the spin.

To assess to which extent the time evolution depends on the
spin state we consider the
difference between the evolution of an $\uparrow$
and of a $\downarrow$ state. We define for $n$ even
\begin{subequations}
\begin{eqnarray}
\widetilde{R}_\uparrow &:= &
\langle \uparrow| \widetilde{R} |\uparrow\rangle_\text{spin}\\
\widetilde{R}_\downarrow &:= &
\langle \downarrow| \widetilde{R}|\downarrow\rangle_\text{spin}
\end{eqnarray}
\end{subequations}
and for $n$ odd
\begin{subequations}
\begin{eqnarray}
\widetilde{R}_\uparrow &:= &
-i\langle \downarrow| \widetilde{R} |\uparrow\rangle_\text{spin}\\
\widetilde{R}_\downarrow &:= &
i\langle \uparrow| \widetilde{R} |\downarrow\rangle_\text{spin}
\end{eqnarray}
\end{subequations}
where the subscript $_\text{spin}$ signifies that we compute
the expectation value only with respect to the Hilbert space of the 
spin. The bosonic operators remain unaltered. Then we
consider 
\begin{subequations}
\begin{eqnarray}
\label{eq:difference-u}
\Delta(t) &:=& \widetilde{R}_\uparrow - \widetilde{R}_\downarrow\\
&=& e^{-iH^\text{eff}t}\; e^{-i\varphi_n(t)}\left[
e^{\Delta_nK}- e^{-\Delta_nK}\right]
\end{eqnarray}
\end{subequations}
as proposed by Lee et al.\ \cite{lee08a}. From the last
formula and \eqref{eq:suppression-general-b} it is obvious that the influence 
of the spin state is small for general sets $\{\omega_i,\lambda_i\}$ if and 
only if $y_n(z)$ is small. Quantitatively, one has
\begin{equation}
\label{eq:result-spin-boson}
y_n(z) = {\cal O}(z^{n+1}) \quad \Leftrightarrow \quad
\Delta(t) = {\cal O}(t^{n+1}) .
\end{equation}
Thereby, we have shown explicitly that the condition 
$y_n(z) = {\cal O}(z^{n+1})$ implies generally that the
coupling between any spin state, i.e., any state of the quantum bit, and 
the bosonic bath is efficiently suppressed if the pulse
sequence obeys \eqref{eq:opt-seq}. Note that this holds
for all choices of $\{\omega_i,\lambda_i\}$.

\section{General Quantum Bath}
\label{sec:quant-sys} 

So far we considered the spin-boson model \eqref{eq:hamilton}. One might
think that the optimized sequence \eqref{eq:opt-seq} is useful only for
this model \cite{khodj07}. This is not the case. 

The first evidence for the general
applicability of \eqref{eq:opt-seq} is the fact that classical
Gaussian noise can equally well be suppressed, see Subsect.\ 
\ref{ss:classical}. Conventional wisdom has it that any
generic model with fluctuations will display Gaussian fluctuations in its
high temperature limit. If this is true the optimized sequence
\eqref{eq:opt-seq} is applicable generally for high temperatures.
Note that the ``high'' temperatures need not be really high. The
inter-spin coupling of nuclear spins is so low that already 1 Kelvin
is sufficient to put a system of nuclear spins at high temperatures.

The second evidence was found by Lee et al.\ \cite{lee08a}. They 
observed analytically for up to $n=9$
that an expansion of $\Delta(t)$ in powers of $t$ 
for a general model yields vanishing coefficients
for the optimized sequence \eqref{eq:opt-seq}. On the basis of
this observation they conjectured that the optimized sequence 
\eqref{eq:opt-seq} is generally applicable for the generic model
for phase decoherence, also called single axis decoherence model
\begin{equation}
\label{eq:hamiltonian-general}
H = \sigma_z A_1 + A_0,
\end{equation}
where $A_0$ and $A_1$ contain only operators from the bath.
Below we use the notation $H_\pm=\pm\sigma_z A_1 + A_0$.

This model does not include spin flips; hence it implies an 
infinite life time $T_1$ as a completely general decoherence model
would do. But the phase decoherence of a precessing
spin in the $xy$ plane is described in full generality because
we do not specify for which operator $A_1$ stands and the bath
dynamics is fully unspecified. It is
described by $A_0$. Such a model is experimentally 
very well justified as the effective model in the limit of 
a large applied magnetic field which implies that  other
couplings between the quantum bit spin and the bath are
averaged out, see for instance Refs.\ \onlinecite{li07} 
and \onlinecite{li08a}.

We investigate the time evolution $\widetilde{R}$ from $0$ to $t$ with 
$\pi$ pulses at the instants $\delta_j t$ where $j\in\{1,2,\ldots,n\}$.
The $\pi$ pulses are assumed to be ideal; they are given by $\sigma_y$
so that $\widetilde{R}$ is given again by \eqref{eq:tildeR}. Next, 
using \eqref{eq:pauli1}, we shift all the factors $\sigma_y$ to the very 
left side  yielding
\begin{eqnarray}
\nonumber
\widetilde{R} &=& 
\left\{\!\!
\begin{array}{c}
1\\
\sigma_y
\end{array}
\!\!
\right\}
e^{-iH_{(-1)^n}(\delta_{n+1}-\delta_n)t}\,
e^{-iH_{(-1)^{n-1}}(\delta_{n}-\delta_{n-1})t}
\ldots\\
&&\ldots  e^{-iH_-(\delta_{2}-\delta_{1})t} \
e^{-iH_+(\delta_{1}-\delta_{0})t}
\label{eq:tildeR-general}
\end{eqnarray}
where the upper entry between curly brackets refers to an even number 
$n$ of pulses and the lower one to an odd number. 

We define the unitary operators $U_p$ as the product of 
the $p+1$ rightmost factors  on the right side of Eq.\
\eqref{eq:tildeR-general}, that means for $0\leq p\leq n$
\begin{eqnarray}
\nonumber
U_p(t) &:=& 
e^{-iH_{(-1)^p}(\delta_{p+1}-\delta_p)t}\,
e^{-iH_{(-1)^{p-1}}(\delta_{p}-\delta_{p-1})t}
\ldots\\
&&\ldots  e^{-iH_-(\delta_{2}-\delta_{1})t} \
e^{-iH_+(\delta_{1}-\delta_{0})t}.
\end{eqnarray}
This operator can be expanded in a Taylor expansion
with coefficients $ C^{\underline{m}}_p$
\begin{equation}
U_p(t) = \sum_{j=0}^\infty (-it)^j \!\! \sum_{\underline{m}\in B_j}
\sigma_z^{|\underline{m}|} C^{\underline{m}}_p
A_{m_{j}}A_{m_{j-1}}\ldots A_{m_{2}}A_{m_{1}}.
\end{equation}
The set $B_j$ contains all binary words $\underline{m}$
with $j$ letters, i.e., $m_i\in \{0,1\}$ where $m_i$ is
the $i$th letter, $i\in\{ 1,2,\ldots,j\}$. Note that also
leading zeros count. We use
$|\underline{m}|$ for the checksum of $\underline{m}$,
i.e., the sum over all letters $|\underline{m}| :=\sum_{i=1}^j m_i$.
The number of letters  $j$ of $\underline{m}$ shall be denoted
by $||\underline{m}||$. Using $B$ as the union of all $B_j$ with
$j\geq0$ we may denote the expansion by
\begin{equation}
\label{eq:gen-expan}
U_p(t) = \sum_{\underline{m}\in B}(-it)^{||\underline{m}||}
\sigma_z^{|\underline{m}|} C^{\underline{m}}_p
A_{m_{||\underline{m}||}}\ldots A_{m_{2}}A_{m_{1}}.
\end{equation}

Obviously, the coefficients which matter in the end are
those for $p=n$. The statement $\Delta(t)={\cal O}(t^{n+1})$
corresponding to \eqref{eq:result-spin-boson} for the spin-boson model
is equivalent to the vanishing of all the coeffients which are
prefactors of terms depending on the spin state. 
This means that all $C^{\underline{m}}_n$ with
$|\underline{m}|$ \emph{odd} have to vanish as long as 
$n\geq ||\underline{m}||$.

So far no general proof is available that these coefficients vanish
for the sequence \eqref{eq:opt-seq}. But for finite $n$ 
the calculation can be done explicitly by computer algebra. Lee et
al.\ carried out such a calculation up to $n=9$ \cite{lee08a}. We succeeded in
reaching $n=14$ by the help of the following recursion.

Clearly, we know from the expansion of a single exponential that
\begin{equation}
C^{\underline{m}}_0 = \frac{1}{||\underline{m}||!} 
(\delta_1-\delta_0)^{||\underline{m}||}.
\end{equation}
This serves as starting point of our recursion which relies on
\begin{subequations}
\begin{eqnarray}
\nonumber
&&U_{p+1}(t) = e^{-iH_{(-1)^{p+1}}(\delta_{p+2}-\delta_{p+1})t}\  U_{p}(t)
\\ 
\nonumber
&&= \sum_{\underline{w}\in B} \Big\{ (-it)^{||\underline{w}||}
[(-1)^{p+1}\sigma_z]^{|\underline{w}|} 
\frac{(\delta_{p+2}-\delta_{p+1})^{||\underline{w}||}}{||\underline{w}||!} 
\\
&& \hspace{3cm} \cdot A_{w_{||\underline{w}||}}\ldots A_{w_{2}}A_{w_{1}}\Big\}
\\
&&\times
\sum_{\underline{m}\in B}(-it)^{||\underline{m}||}
\sigma_z^{|\underline{m}|} C^{\underline{m}}_p
A_{m_{||\underline{m}||}}\ldots A_{m_{2}}A_{m_{1}}.
\end{eqnarray}
\end{subequations}
The comparison of the arising coefficients with those
in \eqref{eq:gen-expan} leads to the recursion relation
\begin{equation}
\label{eq:genrecurs}
C^{\underline{v}}_{p+1} =
\sum_{(\underline{w},\underline{m})=\underline{v}}
\frac{(-1)^{(p+1)|\underline{w}|}}{||\underline{w}||!} 
(\delta_{p+2}-\delta_{p+1})^{||\underline{w}||}
C^{\underline{m}}_p,
\end{equation}
where the sum over $(\underline{w},\underline{m})=\underline{v}$
means that all splittings of the binary word $\underline{v}$ in two 
subwords $\underline{w}$ for the first part and $\underline{m}$
for the second part are considered. Given $\underline{v}$ with
$||\underline{v}||$ letters there are $||\underline{v}||+1$ such
splittings.

The recursion \eqref{eq:genrecurs} can be easily implemented
in computer algebra programmes such as MAPLE. With about 2 Gigabyte
RAM the verification of the vanishing
of the $C^{\underline{v}}_n$ with
odd checksum $|\underline{v}|$ for the optimum
sequence \eqref{eq:opt-seq} up to the order $n=14$ was feasible.
Nevertheless, a general mathematical proof would be highly
desirable.

\section{Influence of the High-Energy Cutoff}
\label{sec:uv} 

Lee et al.\ \cite{lee08a} observed
that the optimized sequence \eqref{eq:opt-seq}, henceforth abbreviated UDD, 
works very well in numerical simulations for GaAs quantum dots
where it does better than the concatenated sequence (CDD) proposed
by Khodjasteh and Lidar \cite{khodj05,khodj07}.\footnote{It should be mentioned
that the CDD was originally proposed for the general decoherence model where
all Pauli matrices of the qubit are coupled to the bath.
In the present article, we consider the special case 
of CDD for single axis models for phase decoherence.} But they found
that qubits made from phosphorous impurities in silicon are better
dynamically decoupled by the CDD sequence. They relate this
result to the applicability of an expansion in time.
Their model consists of a qubit coupled to a spin bath so that
a direct applicability of results obtained for the spin-boson model
is not possible. Yet the question is intriguing whether one can
mimic the qualitative aspects of the spin bath by a bosonic bath.

From the way the general single axis model is treated to derive the
effect of the UDD sequence, see previous section, it is clear that
the expansion in powers of $t$ plays the crucial role. If such 
an expansion in time does not work, for instance because the resulting
expansion is only asymptotically valid, there is no justification
to use the UDD sequence.

The analytically accessible spin-boson model allows us to
investigate the question of the expansion in time in a concrete
example. Inspecting Eqs.\ 
(\ref{eq:finres},\ref{eq:finchi-b},\ref{eq:phi-general-final-d})
one realizes that the existence of the expansion of the signal $s(t)$ depends
on the existence of the expansions of $\chi_n(t)$ and $\varphi_n(t)$.
In order that $\chi_n(t)={\cal O}(t^{n+1})$ the first $n$ derivatives
of  $\chi_n(t)$ must exist and vanish and the $n+1$st derivative
must exist. From \eqref{eq:finchi-b} we see that the expansion
of $\chi_n(t)$ in powers of $t$ is directly related to the expansion
of $y_n(z)$ as in \eqref{eq:def-yn} in powers of $z$ since $z=\omega t$. 
In Subsect.\ \ref{ss:optimization} and in Ref.\ \onlinecite{uhrig07}
we considered only the existence and the vanishing of the
derivatives of $y_n(z)$. The existence of the integral over
the frequencies is no issue as long as a completely hard cutoff
at $\omega_\text{D}$ is considered
\begin{equation}
J_\infty(\omega):=2\alpha\omega\Theta(\omega_\text{D}-\omega)
\end{equation}
for which no ultraviolet (UV) divergence can appear. Hence
all derivatives with respect to time exist for $\chi_n(t)$ and for
$\varphi_n(t)$. This remains also true if the UV cutoff is exponential.

But the physical systems might be such that the UV cutoff is soft
because the spectral density displays power law behavior. We consider
\begin{equation}
\label{eq:def-jgam}
J_\gamma(\omega):=\frac{2\alpha\omega}{1+(\omega/\omega_\text{D})^\gamma}
\end{equation}
as generic form for this situation. Note that $\gamma=\infty$ amounts
up to the completely hard cutoff. The vanishing of the first
$n$ derivatives of $y_n(z)$ implies $y_n(z)=A (\omega t)^{n+1}$ plus higher
terms. But in order to be able to conclude that
$\chi_n(t) = C t^{2n+2}$ plus higher terms the integral
\begin{equation}
C = \frac{A^2}{4} \int_0^\infty  \omega^{2n} J(\omega)
\coth(\beta\omega/2)d\omega
\end{equation}
must exist, i.e., converge. For $J_\gamma(\omega)$ this strictly requires
\begin{equation}
\label{eq:gamma-condition}
\gamma > 2n+2.
\end{equation}

The equivalent consideration for the phase $\varphi_n(t)$ leads
to a less strict condition. If $y_n(z)=A (\omega t)^{n+1}$ plus higher
terms one has $\varphi_n(t) = D t^{n+1}$ which contributes 
the same order $t^{2n+2}$ as $\exp(-2\chi_n)$ to $1-s(t)$ because of the 
cosine in which it appears, see \eqref{eq:finres}. The coefficient $D$
is given by the integral
\begin{equation}
D = \frac{-\Im A}{2} \int_0^\infty  \omega^{n-1} J(\omega)d\omega.
\end{equation}
Its existence requires only
\begin{equation}
\label{eq:gamma-condition2}
\gamma > n+1
\end{equation}
for $J_\gamma(\omega)$. Hence we conclude
that
the condition for the smallness of the deviations resulting from 
$\chi_n$ implies
the condition  for the smallness of the deviations resulting from $\varphi_n$.
For this reason, we will focus on the condition for the smallness
of the exponential suppression by $\chi_n$.

For practical purposes, the existence or non-existence of
certain derivatives is not the ultimate criterion. So below we compare
the effect of various pulse sequences on the signal $s(t)$.
First, we look at the sequence $\text{UDD}_n(t)$ which
is  characterized by \eqref{eq:opt-seq}. It leads
 via \eqref{eq:def-yn}  for an even number of pulses $n$ to
\begin{eqnarray}
\nonumber
y_n^\text{UDD}(z) &=& -2ie^{iz/2}\Bigg\{
\sin\left(\frac{z}{2}\right)+\\
&&2\sum_{j=1}^{n/2}(-1)^j
\sin\left(\frac{z}{2}\cos\left(\frac{j\pi}{n+1}\right)\right)
\Bigg\}.
\end{eqnarray}
Recall Eq.\ \eqref{eq:result-spin-boson} stating the order 
$y_n^\text{UDD}(z) = {\cal O}(z^{n+1})$.

Second, we 
consider the concatenated sequence (CDD) \cite{khodj05,khodj07}.
The zeroth level $\text{CDD}_0(t)$  is the evolution without pulse. Higher
levels are defined recursively by
\begin{subequations}
\begin{eqnarray}
\label{eq:cdd}
\text{CDD}_{l+1}(t) &=& \text{CDD}_{l}(\frac{t}{2}) \circ 
\text{CDD}_{l}(\frac{t}{2})
\ \forall\, l \ \text{odd} \\
\text{CDD}_{l+1}(t) &=& \text{CDD}_{l}(\frac{t}{2}) \circ \Pi \circ 
\text{CDD}_{l}(\frac{t}{2})
\ \forall\, l \ \text{even},\qquad
\end{eqnarray}
\end{subequations}
where $\circ$ stands for the concatenation and $\Pi$ for 
a $\pi$ pulse.
We obtain for the CDD sequence
\begin{equation}
\label{eq:cdd-order}
y_l^\text{CDD}(z) = (-2i)^{l+1}e^{iz/2}\sin(2^{-l-1}z)\prod_{k=1}^l
\sin(2^{-k-1}z),
\end{equation}
where $l$ now stands for the level which is exponentially related to the
number of pulses $n\approx 2^l$. From \eqref{eq:cdd-order} it is easy to see
that $y_l^\text{CDD}(z) = {\cal O}(z^{l+1})$ holds.

Third, we consider the first suggestion \cite{viola98,ban98}, 
namely the periodic bang-bang (BB) control with $n$ pulses and
\begin{equation}
\delta_j = j/(n+1)
\end{equation}
implying (for even $n$)
\begin{equation}
y_n^\text{BB}(z) = -2 i e^{iz/2}\cos\left({z}/{2}\right)
\tan\left({z}/{(2n+2)}\right).
\end{equation}
From this equation one learns $y_n^\text{BB}(z) = {\cal O}(z)$.

Fourth, we consider the Carr-Purcell-Meiboom-Gill (CPMG) sequence
\cite{carr54,meibo58,haebe76}.
This sequence results from the $k$-fold iteration of a two-pulse
cycle of length $\tau=t/k$. The pulses occur at $\tau/4$ and
$3\tau/4$. This cycle corresponds in fact to $\text{UDD}_2(\tau)$
\cite{uhrig07}.
We will come back later to iterations of UDD sequences.
Here we state that CPMG is characterized by
\begin{equation}
\delta_j = (j-1/2)/n
\end{equation}
implying (for even $n$)
\begin{equation}
y_n^\text{CPMG}(z) = 4ie^{iz/2}\sin^2(z/(4n))\frac{\sin(z/2)}{\cos(z/(2n))}.
\end{equation}
From this equation it is clear that  $y_n^\text{CPMG}(z) = {\cal O}(z^3)$.

\begin{figure}[htbp]
    \begin{center}
     \includegraphics[width=0.98\columnwidth,clip]{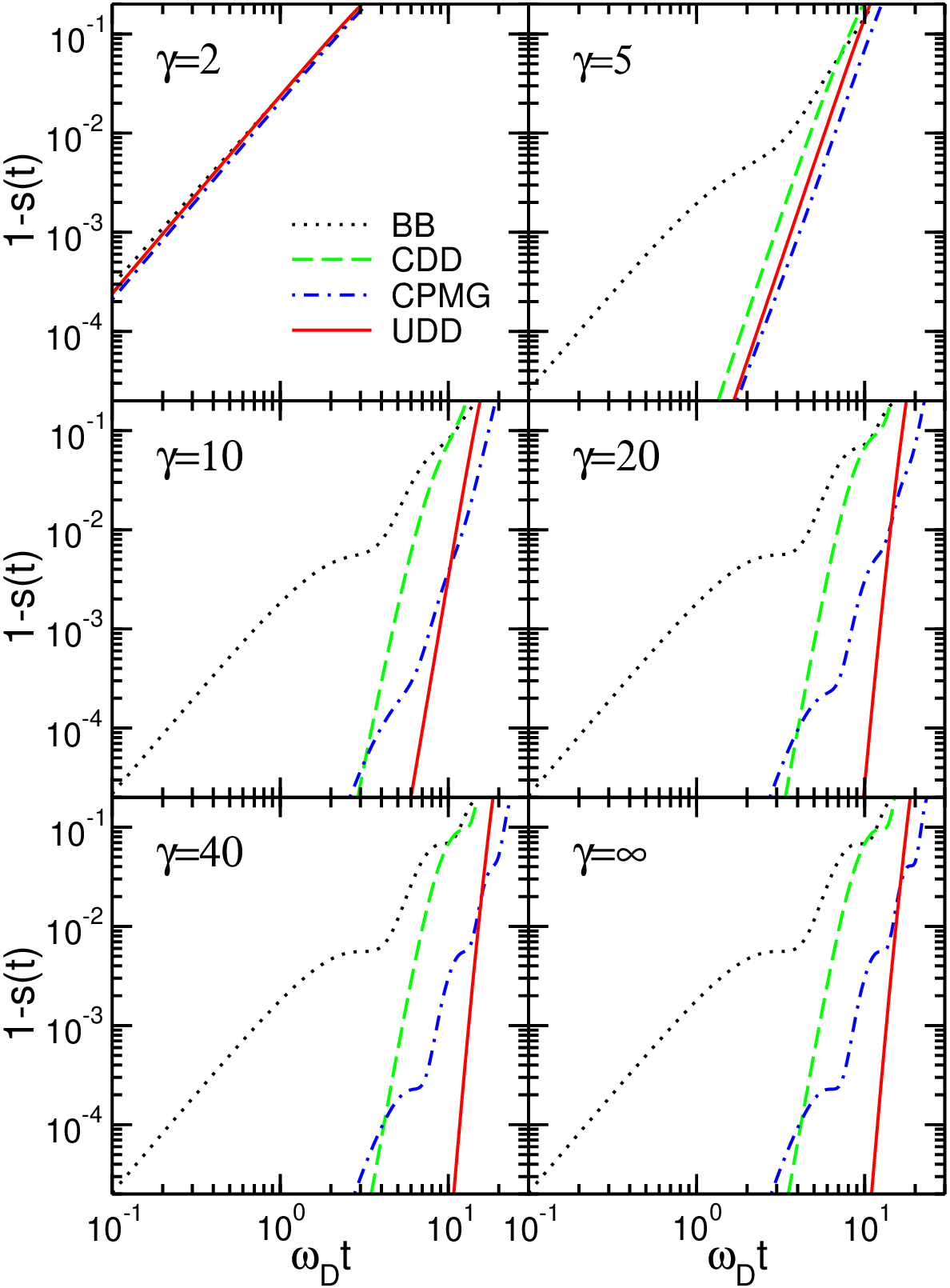}
    \end{center}
    \caption{Various pulse sequences (see main text) are
      compared for various values of the cutoff parameter $\gamma$ in 
      $J_\gamma(\omega)$, see \eqref{eq:def-jgam}. All sequences comprise
      $n=10$ pulses; this values is chosen for better comparison
      because level $l=4$ of the CDD sequence has 10 pulses. The coupling
      value $\alpha$ in the spectral densities $J_\gamma(\omega)$ is fixed
      to $1/4$ and the temperature is zero.}
    \label{fig:various}
\end{figure}
In Fig.\ \ref{fig:various}, the four sequences are compared for
10 $\pi$ pulses at a fixed value $\alpha=1/4$ of the coupling to the bath.
The results for other values of $\alpha$ are very similar.
Furthermore, the temperature is fixed to $T=0$ because the precise
value of the temperature matters only for small frequencies
$\omega\to 0$ while we focus here on high frequencies and the UV
cutoff.

In all six panels it is obvious that the bang-bang (BB) sequence does
worst in accordance with the power law which is only linear. This inefficient
suppression of decoherence also implies that phase effects in 
the signal $s(t)$ due to $\varphi_n$ in Eq.\ \eqref{eq:finres}) 
are seen most strongly leading to the bumps in Fig.\ \ref{fig:various}.
We conclude that one should always try to use one of the other sequences.

Comparing the CDD and the CMPG sequences the CPMG sequence is almost
everywhere advantageous. Only for very low deviations $1-s(t)$ the 
CDD does better because its curve is steeper reflecting a higher
order in $t$: $y_{l=4}^\text{CDD}={\cal O}(t^5)$ while 
$y_n^\text{CPMG}={\cal O}(t^3)$.

Comparing the CDD and the UDD sequences the UDD sequence yields
always lower deviations, except for very soft cutoffs ($\gamma=2$) where
both sequences behave equally. We conclude that we cannot
explain the behavior found by Lee et al. \cite{lee08a} 
where the CDD sequence seemed to outperform the UDD on the basis
of the spin-boson model. Note that the slope of both sequences
in Fig.\ \ref{fig:various} seems to be similar though this
is difficult to tell from the depicted range of parameters. 
But the analytic results clearly states 
$y_4^\text{CDD}(z) = {\cal O}(z^{5})$ while
 $y_{10}^\text{UDD}(z) = {\cal O}(z^{11})$ for
 the same number of pulses, namely $n=10$.

The interesting issue is a comparison of the CPMG and the UDD
sequence. For very soft cutoffs, i.e., low values of $\gamma$,
the CPMG sequence is slightly better. This was also observed
in a model of classical Gaussian noise \cite{cywin08a}.
The UDD sequence, however, performs better for large values of $\gamma$.
Indeed, this finding supports our analytical
condition \eqref{eq:gamma-condition}. As long as $\gamma \lessapprox 2n$
the CPMG sequence with its relatively low order $t^3$ (in $y_n(\omega t)$) does
slightly better than the high-order UDD with $t^{11}$.
But for $\gamma \gtrapprox 2n$ the UDD outperforms the CPMG, especially
for low deviations $1-s(t)$ which matter most for quantum information
processing.

\begin{figure}[htbp]
    \begin{center}
     \includegraphics[width=0.98\columnwidth,clip]{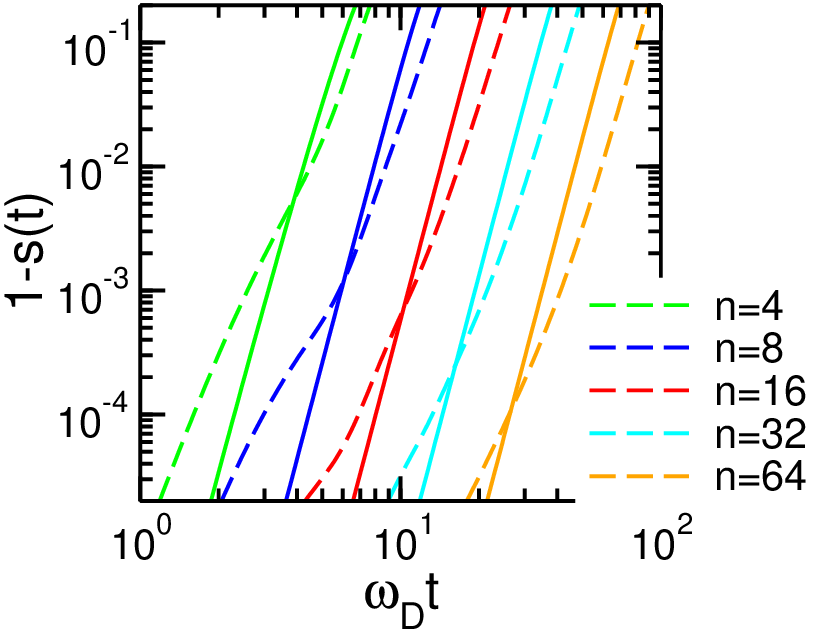}
    \end{center}
    \caption{Comparison of the performance of the CPMG (dashed lines)
      and the UDD sequence (solid lines)
      for various number of pulses $n$ (legend holds for dashed and
      solid lines) and a 
      spectral density with $\gamma=8$ corresponding to a cutoff of
      intermediate hardness. Other values are fixed: $\alpha=1/4, T=0$}
    \label{fig:cpmg-udd}
\end{figure}
We substantiate the comparison between the UDD and the
CPMG sequence further by Fig.\ \ref{fig:cpmg-udd}. The results
go into the same direction as before. As long as $n \lessapprox \gamma/2$
the UDD does significantly better than the CPMG. For
$n\approx \gamma$ the UDD does better than the CPMG at low values
of $1-s(t)\approx 10^{-4}$ while the CPMG is advantageous at 
higher values $1-s(t)\approx 10^{-1}$. For $n> \gamma$,
the CPMG does slightly better than the UDD except for very small values
of  $1-s(t)$. This constitutes a  clear message for applications.

One may wonder whether there is a way to combine the advantages
of the UDD and of the CPMG sequence. Indeed, this is possible by
resorting to hybrid solutions proposed already earlier \cite{uhrig07,dhar06}.
The UDD cycles with low values of $n$ can be iterated. We denote
such a sequence by $\text{iUDD}_{m,c}(t)$ where $m$ stands for 
the number of pulses within one cycle and $c$ for the number
of cycles so that $n=mc$ is the total number of pulses.
This means we consider the concatenation
\begin{equation}
\text{iUDD}_{m,c}(t)
=\left(\text{UDD}_{m}(t/c)\right)^c.
\end{equation}

\begin{figure}[htbp]
    \begin{center}
     \includegraphics[width=0.8\columnwidth,clip]{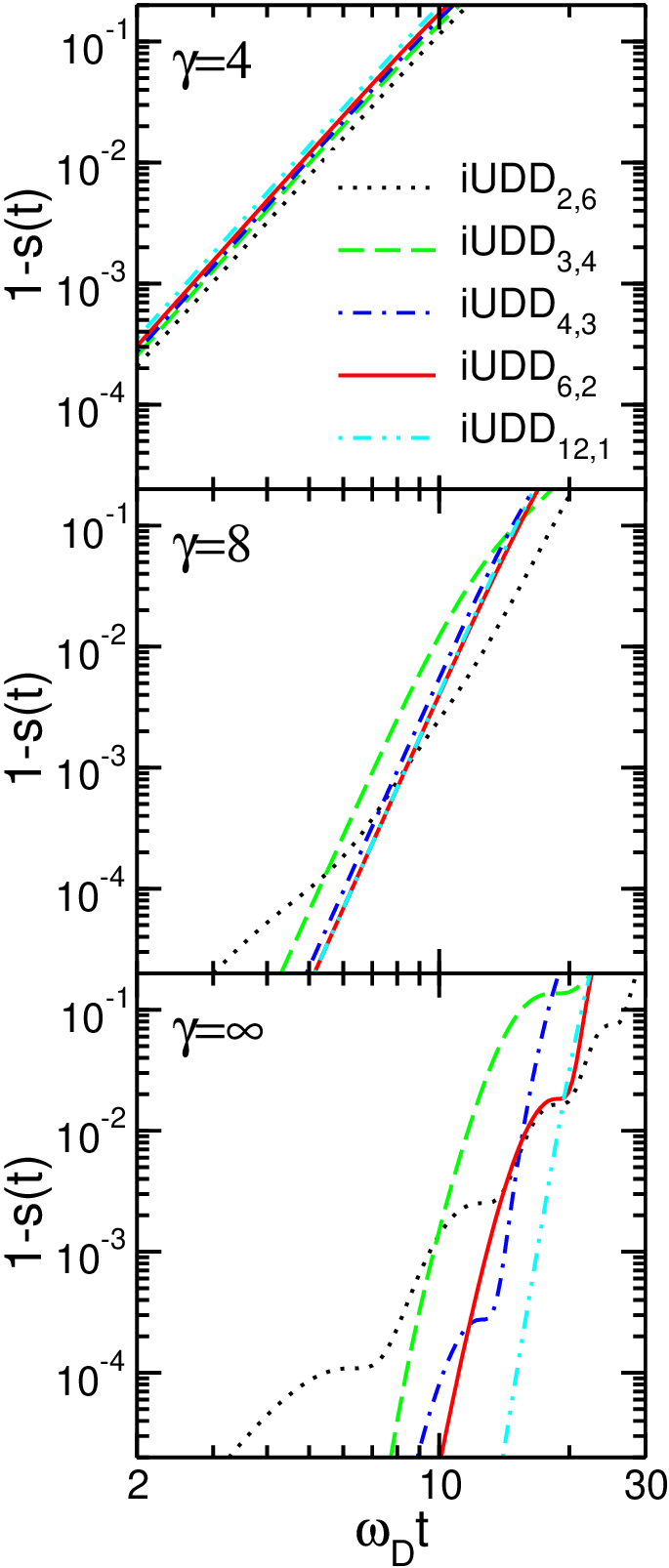}
    \end{center}
    \caption{Comparison of the performance of various iterated 
      $\text{iUDD}_{m,c}(t)$ sequences for a total number of $n=12$
      pulses and various cutoff exponents $\gamma$.
      Other values are $\alpha=1/4, T=0$.}
    \label{fig:iudd}
\end{figure}
A quantitative comparison for iterated iUDD sequences is given
in Fig.\  \ref{fig:iudd} for a total of 12 $\pi$ pulses. Note that
$\text{iUDD}_{2,6}$ is equivalent to CPMG while $\text{iUDD}_{12,1}$
is equivalent to the UDD sequence. The guideline here is the
corollary
\begin{equation}
\label{eq:gamma-condition3}
\gamma > 2m+2
\end{equation}
of \eqref{eq:gamma-condition} where $m$ is the number of pules
within one cycle. It results from the observation that
$y^\text{iUDD}_{m,c}(z)$ is of order $z^{m+1}$ independent from the
number of cycles.

If the condition \eqref{eq:gamma-condition3} is not valid the
use of any sequence of higher order does not pay. This is clearly
seen in the uppermost panel for $\gamma=4$ (very soft cutoff)
in Fig.\  \ref{fig:iudd}. All curves
are almost on top of each other. The CPMG, i.e., $\text{iUDD}_{2,6}$,
is slightly better than the other pulse sequences. 

In the middle panel for $\gamma=8$ (intermediate cutoff)
in Fig.\  \ref{fig:iudd} the situation
has changed. For low deviations $1-s(t)$ the use of the
$\text{iUDD}_{3,4}$ or the $\text{iUDD}_{4,3}$ sequence pays while
the implementation of a larger value of $m$ hardly pays.

In the lowermost panel for $\gamma=\infty$ (hard cutoff) 
in Fig.\  \ref{fig:iudd} the
implementation of higher order sequences is always useful for
low values of $1-s(t)$ as was to be expected.

Fig.\ \ref{fig:iudd}
illustrates that one can gain considerably in coherence without implementing
the fully optimized pulse sequence \eqref{eq:opt-seq}.
Already the implementation of periodic cycles with a moderate number
of pulses can be very helpful. In practice, this strategy
is generally much easier to realize since not so many
special instants in time need to be fine-tuned.

Another remark for experimental realizations is in order. If the pulses
are not ideally tailored then the advantageous of dynamical decoupling
will be thwarted by accumulated pulse errors. So in practice one 
always will be faced with the need to find the optimum tradeoff.
Note, however, that this fact makes it particularly interesting
to reach an optimum suppression of decoherence with a \emph{small}
number of pulses.

\section{Conclusions}
\label{sec:conclusio}

We investigated the suppression of decoherence by sequences of
ideal, instantaneous $\pi$ pulses. The model under study is a spin-boson
model valid for pure dephasing, i.e., for a finite $T_2$ but an
infinite $T_1$. But also the most general  model for phase decoherence
(single-axis decoherence) is considered.

 First, we have provided the
detailed derivation of the equations which were used in our previous
Letter in Ref.\ \onlinecite{uhrig07}. In particular it was rigorously shown
that the sequence \eqref{eq:opt-seq} (UDD) makes the first $n$
derivatives vanish. Furthermore, it was shown that the
results transfer also to the classical case of Gaussian fluctuations.

Second, it was shown that the UDD sequence is advantageous
for any initial state. This important finding was achieved 
by analyzing the corresponding time evolution operator.

Third, we considered the most general model for phase decoherence
 and extended the analytical results of Lee et al. to the
14th order in the time. This was achieved on the basis
of an efficient recursion scheme suitable for implementation
in a computer algebra programme.

Fourth, we investigated the influence of the high-energy
cutoff in the framework of the single-axis spin-boson model.
We compare various pulse sequences which are currently under
debate, namely the periodic bang-bang sequence (BB), the concatenated
dynamical decoupling (CDD), the well-established Carr-Purcell-Meiboom-Gill
sequence (CPMG) and the general iteration of UDD cycles (iUDD).

The most important observation is that decoherence due to baths with
very soft cutoffs are much more difficult to suppress than 
decoherence due to baths with hard cutoffs. For soft cutoffs, the
simpler sequences (CPMG$=\text{iUDD}_{2,c}$ or $\text{iUDD}_{m,c}$ 
with low values of $m$) are completely sufficient. 
Higher order sequences do not
pay. We established a rule of thumb when the implementation of
a more intricate sequence is appropriate. The number of
pulses $m$ in one cycle should not exceed $\gamma/2$ where $\gamma$
is the exponent of the high-energy power law of the decohering
spectral density $J_\gamma(\omega)$, see Eq.\ \eqref{eq:def-jgam}.

By the above results, we have elucicated the possibilities
of dynamical decoupling. Mathematically, important derivations
are provided. Practically, important guidelines are established under which
conditions which sequences are most appropriate.

\acknowledgments
I like to thank J.\ Cardy, D.\ Lidar, S.\ Pasini, P.\ Karbach, T.\ Fischer, 
J.\ Stolze, D.\ Suter, L.\ Viola, and W.\ Witzel for helpful discussions.

% \bibliographystyle{apsrev}
% \bibliography{../../bibinput/liter10}

\end{document}